\documentstyle[sprocl]{article}

\input{epsfig}

\def\Journal#1#2#3#4{{#1} {\bf #2}, #3 (#4)}

\def\NPB{{\em Nucl. Phys.} B}
\def\PLB{{\em Phys. Lett.}  B}

\def\be{\begin{equation}}
\def\ee{\end{equation}}
\def\bea{\begin{eqnarray}}
\def\eea{\end{eqnarray}}

\newcommand{\as}{\alpha_{\rm s}}

\def\siml{{\ \lower-1.2pt\vbox{\hbox{\rlap{$<$}\lower6pt\vbox{\hbox{$\sim$}}}}\ }}

\def\lQ{\Lambda_{\rm QCD}}

\begin{document}
\title{pNRQCD: CONCEPTS and APPLICATIONS\footnote{Talk given at the Fourth Workshop
on Continuous Advances in QCD, Minneapolis (MN), May 12-14, 2000.}}

\author{NORA BRAMBILLA}

\address{Institut f\"ur Theoretische Physik, Philosophenweg 16, 
\\ D 69120, Heidelberg, Germany\\E-mail: n.brambilla@thphys.uni-heidelberg.de}


\maketitle\abstracts{Heavy quark bound state systems, mesons and 
hybrids,  are discussed in the framework of the QCD effective field theory 
called potential Non-Relativistic QCD.}

\section{The physical systems}
Among all the hadrons, the ones that should be the simplest 
to analyze are those entirely composed by heavy quarks (and gluons).
As it is evident from the spectra (e.g. by comparing the characteristic
energy levels splitting  with the value of the mass of the quarks 
in heavy quarkonium), these systems are nonrelativistic. Thus, they can be described 
in first approximation using a Schr\"odinger equation with a potential 
interaction. This amounts to saying  that the heavy quark bound state is characterized
 by three energy scales, hierarchically ordered by the quark velocity 
$v \ll 1$: the mass $m$ (hard scale), the relative momentum $mv$ (soft scale) and 
the binding energy $mv^2$ (ultrasoft scale). 
These scales typically get mixed in any Feynman diagram  bound 
state calculation and originate technical complications (the same happens in QED
 e.g. for positronium). On the other hand, even in a lattice calculation, it is 
difficult to handle physical systems that possess two very different scales,
say $Q \gg q$, since a really demanding relation should then hold for the lattice 
size $L$ and the lattice step $a$: $L^{-1}\ll q \ll Q\ll a^{-1}$.\par

In QCD, a further conceptual complication arises due to the existence of a nonperturbative scale, 
call it $\lQ$, the scale at which the nonperturbative effects  become 
dominant. For heavy quarks only the hard scale $m$ is surely bigger than $\lQ$ and 
can be treated perturbatively.
Many attempts have been made to properly consider the effect of the 
nonperturbative dynamics  on the heavy quark energy levels. On the one hand, phenomenological
or QCD derived confining potential models have  been used inside a Schr\"odinger equation.
Such a picture suffers from many ambiguities, it is strongly model dependent, it consists in 
a by hand superposition of perturbative and nonperturbative effects, 
and it leaves out the physics connected to the retardation effects,
that's to say the physics at the ultrasoft (US) scale. On the other hand, the contribution 
to the Coulombic energy levels due to local condensates has also been evaluated \cite{volleut}.
This correction is of nonpotential type (it is
 analogue to the Lamb shift effect) and it grows out of control for the excited levels.\par
Summarizing, the existence of different scales and the nontrivial features related to 
perturbative (hard scale) and nonperturbative (low scale) effects and to potential and nonpotential 
contributions, complicate considerably  the description of the heavy quark bound states.
I will show here how it is possible to take advantage  of the existence of this hierarchy 
of widely separated energy scales to construct  QCD effective field theories (EFT) 
with less and less degrees of freedom. This leads ultimately to a field theory 
derived quantum mechanical description of these systems. The corresponding EFT is called 
pNRQCD (potential NonRelativistic QCD) \cite{pNRQCD1,pNRQCD2}. Here, all the dynamical regimes
 are
organized in a systematic expansion. 
 
\section{The QCD effective theories for these systems}
The idea is that since the typical scales of a nonrelativistic bound state system
are widely separated, it is possible to perform an expansion of one scale in terms of 
the others. Roughly speaking, first one expands in ${1/m} $ and then one performs an expansion 
in the inverse of the soft scale, the so called multipole expansion 
\cite{multpol,volleut,pNRQCD1,pNRQCD2}.
The effective theory supplies us with the procedure to make this expansion consistent 
with the ultraviolet behaviour of QCD  and consistent  
with a systematic power counting in the small expansion parameter 
$v$.  As I will explain below, it depends on  the physical system in consideration 
and on the actual position  of  $\lQ$ 
into this hierarchy of scales, if  it is possible or not to take advantage of the further simplification 
of performing the calculations in a perturbative expansion in $\alpha_s$.

\subsection{NRQCD}
First we pass from QCD to NRQCD by integrating out the hard scale $m$. This involves 
performing an expansion in ${1/m}$, which, in the two-fermions sector is
 of the type of the Foldy-Wouthuysen transformation.
Since we are modifying  the ultraviolet behaviour of the theory, matching coefficients and 
new operators have to be added  in order to mock up the effects of heavy particles  and 
high energy modes  into the low energy EFT.
In this case, since the 
scale of the mass of the heavy quark is perturbative, the scale $\mu$
of the matching from QCD to NRQCD, $mv <\mu < m$, lies also in the perturbative regime.
Then, the  hard scale is integrated out  by comparing on shell amplitudes,
order by order in ${1/m}$ and in $\alpha_s$,  in QCD and in NRQCD. The difference is 
encoded into  the matching coefficients that typically depend non-analytically on the scale 
which has been integrated out,  in this case $m$.
We work here in dimensional regularization,
$\overline{MS}$ scheme, and with quark pole masses.
Up to order ${1/m^2}$ the Lagrangian of NRQCD \cite{nrqcd} reads:
\begin{eqnarray}
& &\ L = \psi^\dagger\left(iD_0 +  c_2  {{\bf D}^2\over 2  m} + 
 c_4  {{\bf D}^4\over 8  m^3} 
+  c_F  g { {\bf  S}\cdot {\bf B} \over  m } +  c_D  g { {\bf D}\cdot{\bf E} 
 - {\bf E}\cdot{\bf D} \over 8  m^2}\right. \label{NRQCD}\\ 
& & \left. + i  c_S  g {{\bf S}\cdot({\bf D}\times{\bf E} - 
{\bf E} \times {\bf D}) \over 4  m^2}\right) \psi
+ \hbox{\,  antiquark terms} + \hbox{\, terms with light quarks} \nonumber\\
& & 
-{ b_1 \over 4} F_{\mu\nu}^a F^{a\,\mu\nu}  
+ {b_2\over  m^2} F_{\mu\nu}^a D^2 F^{a\,\mu\nu}
 + { b_3 \over m^2} g  f_{abc} F_{\mu\nu}^a  F_{\mu\alpha}^b F_{\nu\alpha}^c \nonumber \\
& & + { d_1 \over  m^2} \psi^\dagger \psi \chi^\dagger \chi   
+ { d_2  \over  m^2}  \psi^\dagger {\bf S} \psi \chi^\dagger {\bf S} \chi 
+ { d_3 \over  m^2} \psi^\dagger  T^a  \psi \chi^\dagger  T^a  \chi 
+ \!{ d_4 \over  m^2} 
\psi^\dagger  T^a {\bf S}  \psi \chi^\dagger  T^a {\bf  S}  \chi, \nonumber
\end{eqnarray}
$\psi$ and $\chi$ being respectively the quark and the antiquark field; $D^\mu$
is the covariant derivative, ${\bf E}$ and ${\bf B}$ are chromoelectric and chromomagnetic 
fields, $F^{\mu\nu}$ is the gluon field strength, ${\bf S}$ is the total spin,
$T^a$ is the color generator, $g$ is the coupling constant.  
$c_i,b_i,d_i$ are matching coefficients. They depend on $\mu$ and $m$ and 
are known in the literature at a different level of precision. The $\mu$ dependence in the matching 
coefficients cancels against  the $\mu$ dependence of the operators in the 
Lagrangian.
The gluonic part  in the third line comes from the vacuum polarization 
while the last line contains four quark operators.\par  

Many applications of NRQCD have been made both in lattice calculations and in 
the continuum, see e.g. \cite{nrqcd,rev}.
Here, however I just introduced NRQCD as a step towards pNRQCD.
I recall in fact that in NRQCD two dynamical scales are still present, $mv$ and 
$mv^2$. These scales get entangled and  obscure the power counting, i.e.
the matrix elements of the operators in (\ref{NRQCD}) 
do not have a unique power counting  but they also contribute to 
subleading orders in $v$. 
In the next section I will show how it is possible to integrate out also 
the momentum scale, obtaining an effective theory at the ultrasoft scale.
This effective theory is called pNRQCD.

\subsection{pNRQCD}
When we  integrate out the soft scale $mv$, there are two possible situations depending on the 
relative position of $\lQ$ with respect to the other scales.
If $\lQ$ is smaller than $mv$, then also the matching from NRQCD to
pNRQCD may be performed in perturbation theory. If $\lQ \simeq mv$ the 
soft scale is already nonperturbative and the matching to pNRQCD has to be performed 
 avoiding any expansion in $\alpha_s$. Roughly speaking, we can say that the 
lowest excitations of quarkonium belong to the first situation while the excited states 
belong to the second situation. Indeed, the typical radius of the bound state is proportional
to the inverse of the soft scale $r \simeq 1/mv$ and thus for the lowest states 
the condition $mv \ll \lQ$ may be fulfilled. In Fig.1  
I present the typical radius of various mesons against  the curve of a 
phenomenological potential  that displays a nonperturbative (linear) behaviour around 0.2 fm.
In the following  I will discuss both situations.
\vskip -0.9truecm
\begin{figure}[htc]
\vskip -1truecm
\makebox[03cm]{\phantom b}
\epsfxsize=5.9truecm \epsfbox{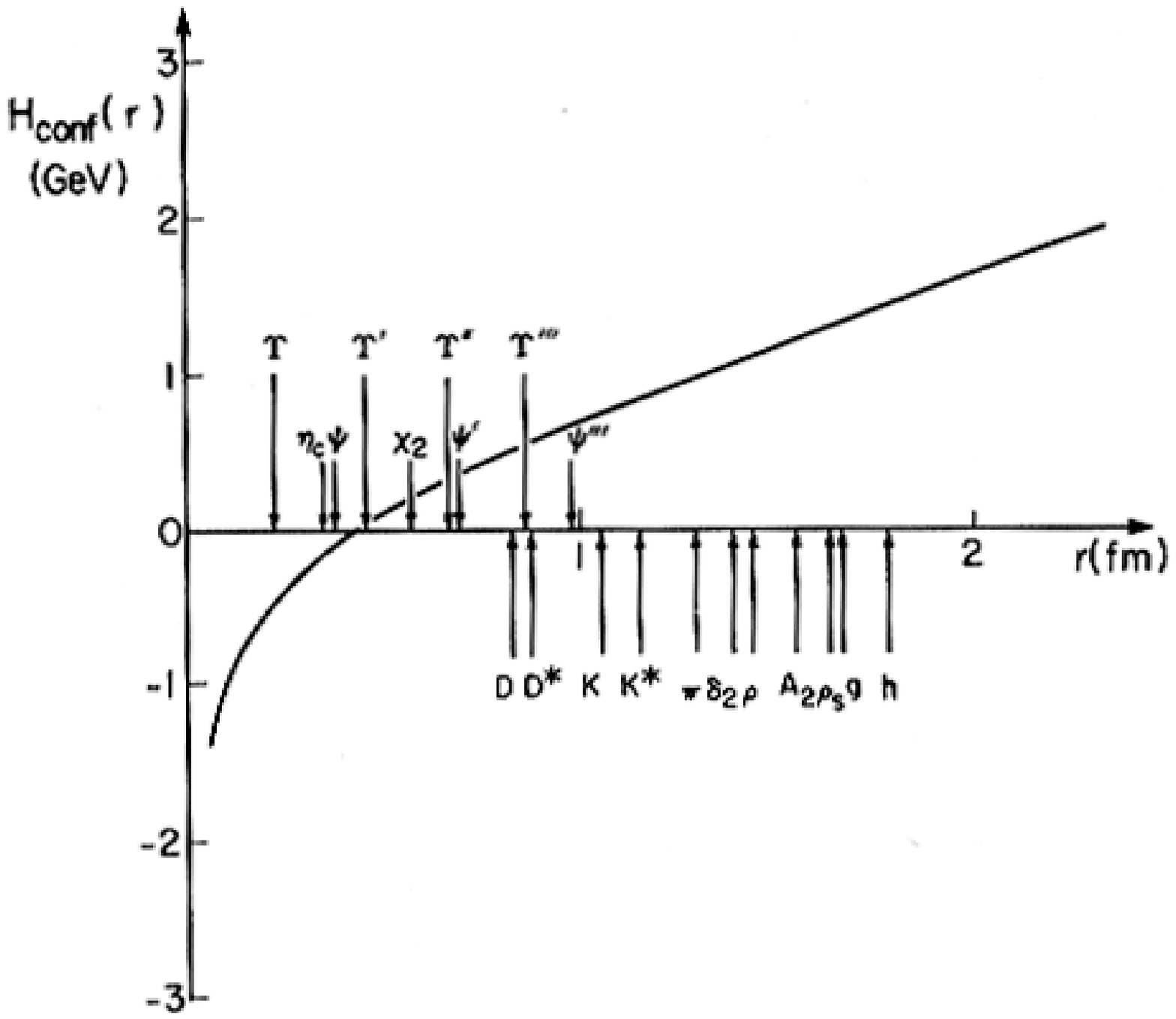}
\vskip -2.4truecm
\caption{ $r_{q\bar{q}}$ of various mesons  vs the Cornell potential, 
from$ ^7$.}
\label{uno}
\end{figure}

\section{pNRQCD for $mv \gg \lQ$}
We denote by ${\bf R}\equiv {({\bf x}_1+{\bf x}_2})/2$ the center of mass of the $Q\bar{Q}$ 
system and 
by  ${\bf r\equiv {\bf x}_1 -{\bf x}_2}$ the relative distance. 
The matching to pNRQCD is perturbative.
 At the scale of the matching $\mu^\prime$ 
($mv \gg \mu^\prime \gg mv^2, \lQ$) we have still quarks and gluons.
The effective degrees of freedom are: $Q\bar{Q}$ states (that can be decomposed into 
a singlet $S({\bf R},{\bf r},t)$ and an octet $O({\bf R},{\bf r},t)$
under color transformations) with energy of order of the next relevant 
scale, $O(\Lambda_{QCD},mv^2)$ and momentum\footnote{Notice  that although,
for simplicity, we described the matching between NRQCD and pNRQCD as integrating out 
the soft scale,  it should be clear that the relative momentum ${\bf p}$ of the quarks is
still soft.}  ${\bf p}$ of order $O(mv)$,  plus 
ultrasoft gluons $A_\mu({\bf R},t)$ with energy 
and momentum of order  $O(\lQ,mv^2)$. Notice that all the  gluon fields are multipole 
expanded. The Lagrangian is then  an expansion 
in the small quantities  $ {p/m}$, ${ 1/r  m}$ and in   
$O(\Lambda_{\rm QCD}, m v^2)\times r$.

\subsection{The pNRQCD Lagrangian}
At the next-to-leading order (NLO) in the multipole expansion,
 i.e. at $O(r)$, we get \cite{pNRQCD1,pNRQCD2} 
\begin{eqnarray}
& & 
 L^{(1)}_{\rm pNRQCD} 
=   -{1\over 4}  F_{\mu\nu}^a F^{\mu\nu\,a} 
+{\rm Tr} \left\{{\rm S}^\dagger \left( i\partial_0 - {{\bf p}^2\over  m } 
- V_s
-\sum_{n=1} { V^{(n)}_s\over  m^n} \right){\rm S} \right\} \label{pnrqcd}\\
& & + {\rm Tr} \left\{ {\rm  O^\dagger} \left( i{ D_0} - {{\bf p}^2\over  m}-V_o 
- \sum_{n=1} { V^{(n)}_o\over  m^n} \right){\rm  O} \right\}\nonumber \\
& & 
+ {g}  V_A{\rm Tr} \left\{  {\rm  O^\dagger} {\bf r}\cdot{ \bf E}\,{\rm  S} 
+ {\rm  S^\dagger} {\bf r}\cdot{ \bf E} \,{\rm  O} \right\} 
+ { g} { V_B\over 2} {\rm Tr} \left\{  {\rm  O^\dagger} 
{ \bf r}\cdot{ \bf E}\,{\rm  O}
+ {\rm  O^\dagger} {\rm  O }{ \bf r}\cdot{ \bf E} \right\}.\nonumber
\end{eqnarray} 
The matching coefficients 
$V_j$ are functions of $\mu, \mu^\prime, {\bf r}, {\bf p}, {\bf S}_1, {\bf S_2}$.
The equivalence of pNRQCD to NRQCD, and hence to QCD, is enforced
by requiring the Green functions of both effective theories to be equal (matching).
In practice, appropriate off shell amplitudes are compared in NRQCD and in pNRQCD,
order by order in the expansion in $1/m$, $\alpha_s$  and in the multipole expansion.
The difference is encoded in these potential-like matching coefficients 
that depend non-analytically on the scale that has been integrated out (in this case ${\bf r})$.
 Recalling that ${\bf r} \simeq 1/mv$ and that the operators count like the next relevant 
scale, $O(mv^2,\lQ)$, to the power of the dimension, it follows that  
each term in  the pNRQCD Lagrangian has a definite power counting.  This feature makes 
$L_{\rm pNRQCD}$ the most suitable tool for a bound state calculation: being interested 
in knowing the energy levels up to some power $v^n$, we just need to evaluate the contributions
of this size in the Lagrangian. 
The singlet sector of the Lagrangian would give rise to equations of motion of the 
Schr\"odinger type, while the last line of (\ref{pnrqcd})
contains  retardation (or nonpotential) effects that 
start at the NLO in the multipole expansion. At this order the nonpotential
effects come from the singlet-octet interaction mediated by a ultrasoft chromoelectric 
field. \par
Summarizing, pNRQCD is equivalent to QCD, it has potential terms, thus embracing potential 
models, and it has ultrasoft gluons 
incorporated in a second-quantized, gauge-invariant and systematic way. 
 Moreover, the power counting in $v$ is explicit  and subsequent corrections 
both in the $1/m$ expansion as in the multipole expansion can be added systematically. 
Hard and soft effects are separated from the ultrasoft (low-energy) effects.

The Feynman rules corresponding to the Lagrangian (\ref{pnrqcd}) are shown in Fig.2. 

At zero order in the multipole expansion the singlet and the octet decouple and their equations 
of motion turn out to be Schr\"odinger-like.  Then, the first question comes: is the singlet 
matching coefficient $V_s$ equal to the static heavy quark potential? 
The answer depends on the actual ratio of $mv^2$ and $\lQ$.
In the EFT language the potential is defined upon 
the integration of all the scales {\it up to the ultrasoft 
scale $mv^2$}.  We can imagine two different  
situations: if $\lQ \siml mv^2$, then $V_s$ is   the static heavy 
quark potential;
if $mv \ll \lQ \ll mv^2$,
 then we have to integrate out also  $\lQ$ in order to get the potential. In this 
 integration the potential acquires short range nonperturbative contributions
 as I will show in Sec.3.3.\par
A second interesting question is what is the relation between  $V_s$ and     
the energy $E_s$ between quark static sources, which is defined as
\begin{equation}
E_s(r) =  \displaystyle\lim_{T\to\infty} {i \over T} \ln  \langle W_\Box \rangle,
\label{eqW}
\end{equation}
being $W_\Box$  the static Wilson loop of size ${\bf r} \times T$,  and 
the symbol $\langle ~~ \rangle$ being the average over the gauge fields.
$E_s$ is  often  used as a static potential inside 
the Schr\"odinger equation, assuming that  the Born-Oppenheimer approximation holds.
We answer these questions  by  performing  explicitly the  singlet
matching  at order $1/m^0$ and at  the NLO in the multipole expansion.  
\begin{figure}[htb]
\makebox[0.5cm]{\phantom b}
\epsfxsize=1.2truecm \epsfbox{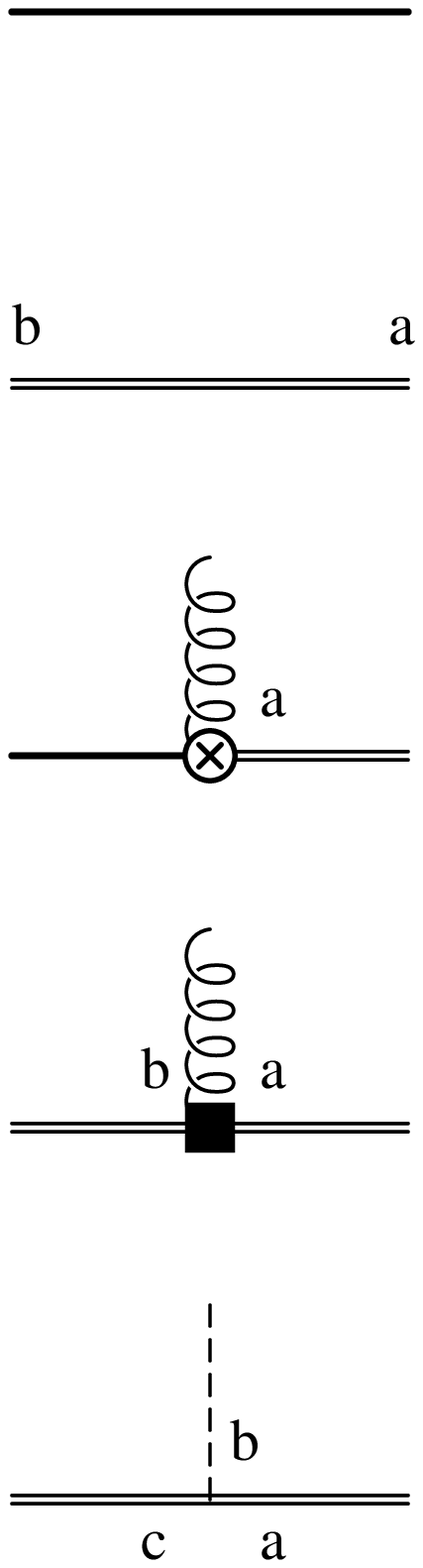}
\put(25,126){\small $= \theta(T) e^{\displaystyle -i V_s T}$}\put(210,126){\small singlet propagator}
\put(25,96){\small $= \theta(T)\Bigg(e^{\displaystyle -i V_o T} 
e^{\displaystyle - i g \int_{-T/2}^{T/2} \!\!\! dt 
                 \, A_0^{\rm adj}}\Bigg)_{ab}$} \put(210,96){\small octet propagator}
\put(25,66){\small $= i g V_A \displaystyle\sqrt{T_F\over N_c} {\bf r} \cdot {\bf E}^a$}
\put(210,66){\small singlet--octet vertex}
\put(25,36){\small$= i g \displaystyle{V_B\over 2} d^{abc} {\bf r} \cdot {\bf E}^c$}
\put(210,36){\small octet--octet vertex}
\put(25,6){\small $=  g f^{abc}$}\put(210,6){\small Coulomb octet--octet vertex}
\caption{ \it Propagators and vertices of the pNRQCD Lagrangian (at order $1/m^0$ and 
at the NLO in the multipole expansion).}
\label{due}
\end{figure}

\subsection{The matching procedure: the singlet ($\lQ \siml mv^2$)}
The matching can be done once the interpolating fields for $S$ and $O^a$ have been identified in NRQCD. 
The former need to have the same quantum numbers and the same transformation properties as the latter. 
The correspondence is not one-to-one. Given an interpolating field in NRQCD, 
there is an infinite number of combinations of singlet and octet wave-functions 
with ultrasoft fields, which have the same quantum numbers 
and, therefore, have a non vanishing overlap with the NRQCD operator. However, the operators in pNRQCD 
can be organized according to the counting of the multipole expansion. 
For instance, for the singlet we have  
\begin{equation}
\chi^\dagger({\bf x}_2,t) \phi({\bf x}_2,{\bf x}_1;t) \psi({\bf x}_1,t) =  Z^{1/2}_s(r) S({\bf R},{\bf r},t) 
+ Z^{1/2}_{E,s}(r) r \, {\bf r}\cdot{\bf E}^a({\bf R},t) O^a({\bf R},{\bf r},t) + \dots,  
\label{Sdef}
\end{equation}
and  for the octet 
\begin{eqnarray}
\!\!\!\chi^\dagger({\bf x}_2,t) \phi({\bf x}_2,{\bf R};t) T^a \phi({\bf R},{\bf x}_1;t) \psi({\bf x}_1,t) 
&=& Z^{1/2}_o(r) O^a({\bf R},{\bf r},t) \label{Odef}\\
&+& Z^{1/2}_{E,o}(r) r \, {\bf r}\cdot{\bf E}^a({\bf R},t) S({\bf R},{\bf r},t) + \dots, 
\nonumber
\end{eqnarray}
$
\phi({\bf y},{\bf x};t)\equiv {\rm P} \exp \{ ig 
\int_0^1 \!\! ds \, ({\bf y} - {\bf x}) \cdot {\bf A}({\bf x} - s({\bf x} - {\bf y}),t) \}$, $Z_i$ being 
normalization factors.
These operators guarantee a
leading overlap with the singlet and the octet wave-functions respectively. Higher order corrections are 
suppressed in the multipole expansion. 
The expressions for the octet can be made manifestly 
gauge-invariant by inserting a magnetic field in place 
of the pure color matrix. The fact that the matching can be done in a completely gauge-invariant 
way enables us to generalize pNRQCD to the case in which $\lQ \simeq mv$, i.e. to the nonperturbative
 matching, cf. Sec. 5.\par 
Now, in order to get the singlet potential, we compare 4-quark Green functions. From (\ref{Sdef}),
we take in NRQCD
\begin{equation}
I = \delta^3({\bf x}_1 - {\bf y}_1) \delta^3({\bf x}_2 - {\bf y}_2) \langle W_\Box \rangle , 
\label{vsnrqcd}
\end{equation}
 and we equate  
(\ref{vsnrqcd}) to the singlet propagator in pNRQCD
at NLO in the multipole expansion (cf. Fig. 3 for a diagrammatic representation)
\begin{eqnarray}
& &
\!\!\!\!\!\!\!\!\!
I = Z_s(r) \delta^3({\bf x}_1 - {\bf y}_1) \delta^3({\bf x}_2 - {\bf y}_2) e^{-iTV_s(r)} \times
\label{vspnrqcdus}\\
& &
\!\!\!\!\!\!\!\!\!\!\!\!\Bigg( 1 -{ g^2 \over N_c} T_F V_A^2 (r)\int_{-T/2}^{T/2} \!\!\! dt 
\int_{-T/2}^{t} \!\!\! dt^\prime e^{-i(t-t^\prime)(V_o-V_s)} 
\langle {\bf r}\cdot {\bf E}^a(t) \phi(t,t^\prime)^{\rm adj}_{ab}
{\bf r}\cdot {\bf E}^b(t^\prime)\rangle \Bigg),
\nonumber
\end{eqnarray}
where $\phi^{\rm adj}$ is a Schwinger (straight-line) string in the adjoint representation 
and fields with only temporal argument are evaluated in the centre-of-mass coordinate.
From here one gets 
the singlet static potential $V_s$ (and the singlet wave-function normalization $Z_s$):
\begin{equation}
V_s(r) = E_s(r)\big\vert_{\rm 2-loop+NNNLL} 
+ C_F {\alpha_{\rm s}\over r} {\alpha^3_{\rm s}\over \pi}
{C_A^3\over 12} \ln {C_A \alpha_{\rm s} \over 2 r \mu^\prime},  
\label{vsu0}
\end{equation}
where $E_s$ has been defined in Eq. (\ref{eqW})\footnote{The non-relativistic limit
described by the Schr\"odinger equation with the static potential is 
called 'leading-order' (LO);  contributions corresponding to
corrections of order $v^n$ to this limit are called ${\rm N}^n{\rm LO}$. LL means 
'leading-log'.  In the perturbative situation $v=\alpha_s$.}
We note that $V_s$ and $E_s$ would coincide 
in QED and that therefore the effect we are studying here is a genuine QCD feature.  
\begin{figure}[htb]
\makebox[3.6cm]{\phantom b}
\epsfxsize=5.5truecm \epsfbox{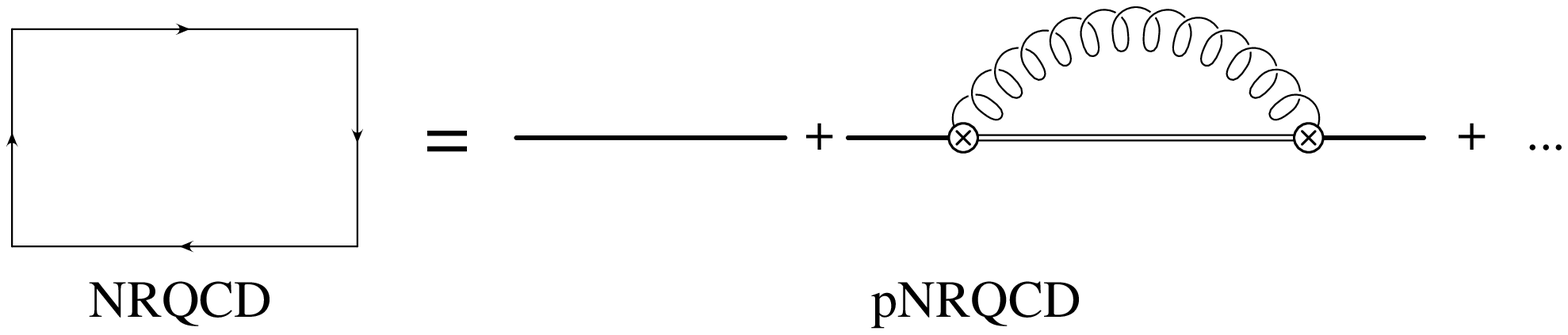}
\vskip -0.4truecm
\caption{The matching between the Wilson loop in NRQCD and the singlet propagator in pNRQCD.}
\vskip -0.2truecm
\end{figure}
The leading log contributions to the NNNLO   
arise from three loops diagrams that are infrared divergent. The final result reads
\begin{eqnarray}
V_s(r,\mu^\prime) &=& - C_F{\alpha_V (r,\mu^\prime)\over  r},\label{vpot} \\
\alpha_V(r,\mu^\prime) &=& \as( r)\Big[1 + \tilde{a}_1  \as(r) 
+ \tilde{a}_2   (\as(r))^{2} + { \as^3 \over \pi} {C_A^3\over 12} \ln {\mu^\prime  r }\Big].
\nonumber
\end{eqnarray}
This is the LL three loop evaluation of the static heavy quark potential 
in the case $\lQ \siml mv^2$. The one loop and two-loop contributions come from \cite{looppot} and 
the three loop LL from \cite{pot}. \par
This situation is expected to hold for toponium and for 
the bottomonium (charmonium?) ground state. 
The explicit $\mu^\prime$ dependence of $V_s$   originates from the fact that the US 
degrees of freedom (which have the same scale of the kinetic energy and therefore do not belong to 
the potential) have been explicitly subtracted out from the static Wilson loop. 
This fact is not surprising if we understand the heavy quarkonium potential 
as a matching coefficient of pNRQCD. As a consequence even in a purely perturbative 
regime the static heavy quarkonium potential (as well as $\alpha_V$) 
turns out to be an infrared sensitive quantity. 
In this  situation nonperturbative effects are only of non-potential nature
and appear in the form of local (\`a la Leutwyler--Voloshin) or 
nonlocal  condensates. When calculating any physical observable, the $\mu^\prime$ 
dependence  cancels against $\mu^\prime$-dependent contributions 
coming from the dynamical US gluons \cite{spect}.

In a similar way, by matching appropriate Green functions, 
we can get  the octet matching potential $V_o$ and the other matching potentials 
\cite{pNRQCD2,spect}.

\subsection{The singlet potential in the situation $ mv \gg \Lambda_{\rm QCD} \gg mv^2$}
Since in this situation there is a physical scale ($\lQ$)
above the US scale, a potential can be properly defined only once this scale has been integrated out.
At the NLO in the multipole expansion we get 
\begin{eqnarray}
& &
\!\!\!\!\!\!\!\!\!\!\!\!\!\!\!V_s(r) = -C_F {\alpha_{V}(r,\mu^\prime)\over  r }
 -i{g^2 \over N_c}T_F V_A^2(r){r^2\over 3} \int_0^\infty\!\!\!\! dt 
e^{-it(V_o-V_s)} 
\langle {{\bf E}^a}(t)\phi(t,0)^{\rm adj}_{ab}{\bf E}^{b}(0) \rangle(\mu^\prime).
\nonumber \\
& & \label{vsnp}
\end{eqnarray}
Therefore, the heavy quarkonium static potential $V_s$ is given in this situation 
by the sum of the purely 
perturbative piece calculated in Eq. (\ref{vpot}) and a new term carrying   
also nonperturbative contributions (contained into non-local gluon field correlators). 
This last one can be organized as a series of power of $r^n$ by expanding 
$\exp\{-it(V_o-V_s)\}=1 -it(V_o-V_s)+ \dots$ (since $t\simeq 1/\lQ$, $V_o-V_s \simeq mv^2$). 
Typically the nonperturbative piece 
of Eq. (\ref{vsnp}) absorbs the $\mu^\prime$ dependence of $\alpha_V$  
so that the resulting potential $V_s$ is now scale independent.\\
We notice that the leading nonperturbative term could be as important as the perturbative potential 
once the power counting is established and, if so, it should be kept exact when 
solving the Schr\"odinger  equation. In Table 1 we summarize the different kinematic situations.
\begin{table}[htb]
\makebox[2.2cm]{\phantom b}
\begin{tabular}{|c|c|l|l|}
\hline
$mv$&$mv^2$&potential&ultrasoft corrections\\\hline
$\gg \Lambda_{\rm QCD}$&$\gg \Lambda_{\rm QCD}$&perturbative&local condensates\\
$\gg \Lambda_{\rm QCD}$&$\sim \Lambda_{\rm QCD}$&perturbative&non-local condensates\\
$\gg \Lambda_{\rm QCD}$&$\ll \Lambda_{\rm QCD}$&perturbative + & No US (if light quarks\\
&$~$&short-range nonpert.& $\,\,\,$ are not considered)\\ 
\hline
\end{tabular}
\caption{Summary of the different kinematic situations.}
\label{tab2}
\vspace{-0.1cm}
\end{table}

We can also consider Eq.(\ref{vsnp}) in the limit of static sources. In this case there are 
no dynamical scales involved and therefore in the short range we have  
$\exp{\{-it (V_o-V_s)\}}
\langle {{\bf E}^a}(t)\phi(t,0)^{\rm adj}_{ab}{\bf E}^{b}(0)\rangle$   
$\simeq$  $\exp{\{-it (V_o-V_s)\}}\langle E(0)^2\rangle$. This gives 
\begin{equation}
E_s(r)\simeq -C_{\rm F}{ \alpha_{\rm V}(r, {\alpha_s\over r})\over r}- {r^3\over 3} {g^2\over 
N_c^2 \alpha_s}\langle E^2(0)\rangle
\label{flo}
\end{equation}
which coincides with \cite{flory}.

\section{Application of pNRQCD ($\lQ < mv$)}
\begin{itemize}
\item{\it Quarkonium spectrum at leading-log of NNNLO}\\
The complete leading-log terms of the NNNLO corrections to pNRQCD have been calculated \cite{spect}.
As a byproduct  the leading logs at $O(m\alpha_s^5)$ in the heavy quarkonium spectrum
 have been obtained.
When $\lQ \ll m\alpha_s^2$, these leading logs  give the complete $O(m\alpha_s^5\ln\alpha_s)$ 
corrections to the heavy quarkonium spectrum (plus the nonpotential contributions coming from 
local condensates) \cite{spect}. The result is important at least 
for $t \bar{t}$ production and $\Upsilon$ physics. In the first case this result is a first step 
towards the goal of reaching  a 100 MeV sensitivity on the top quark mass from the 
$t\bar{t}$ cross section near threshold to be measured at future linear colliders. In the second 
case it will improve our knowledge on the $b$ mass. In both cases the LL contributions of 
NNNLO have been found to be relevant \cite{tt,mb}.
\item{\it Renormalons}\\ 
The infrared sensitivity of the static potential can also be expressed in the 
renormalons language,  i.e. we can say that $V_s$, as defined in Eq.(\ref{vpot}), suffers from IR 
renormalons ambiguities with the following structure
\begin{equation}
V_s(r) \vert_{\rm IR\, ren} = C_0 + C_2 r^2 + \dots
\label{ren1}
\end{equation}
The constant $C_0 \sim \Lambda_{\rm QCD}$ is known to be cancelled by the pole 
mass IR renormalon ($2 m_{\rm pole}\vert_{\rm IR\, ren} = - C_0)$. 
While Eq. (\ref{vsnp}) provides us with the explicit expression for the 
operator which absorbs the $C_2 \sim \Lambda_{\rm QCD}^3$ ambiguity \cite{pNRQCD2}.
The renormalon cancellation issue plays an important role in quarkonium phenomenology
\cite{ren}. 
\item{\it Hybrids and gluelumps}\\ 
pNRQCD gives model independent predictions on the behaviour of the hybrid static potentials.
In particular it predicts for these  potentials an octet behaviour at very short distances 
and it correctly states all the degeneration patterns in the small $r$ region \cite{pNRQCD2}.
Moreover, it allows to relate the mass of the gluelumps to the correlation lengths of some 
nonlocal vacuum field strength correlators \cite{hybrid,pNRQCD2}.
\item{\it Quarkonium scattering, Van Der Waals forces, Quarkonium Production.} 
Work is in progress on these applications.
\item{\it Renormalization group improvement}\\
Up to now, it has been performed at the level of the static potential \cite{rg}.
\end{itemize}

\section{pNRQCD for $\Lambda \simeq mv$}
In this case the potential interaction is dominated by nonperturbative effects.
This is the most interesting situation, in which most of the mesons lie (cf. Fig.1).\par
A large effort has been made in the last decades in order to obtain from QCD
the nonperturbative potentials in the Wilson loop approach \cite{np,mod,rev}.
pNRQCD  allows us to obtain via a nonperturbative matching all the nonperturbative 
potentials \cite{m1}. From this respect I want to discuss few concepts and results.\par
In pNRQCD a  potential picture for heavy quarkonium 
emerges at the leading order in the US expansion 
under the condition that the matching between  NRQCD and pNRQCD  can be performed within 
an expansion in $1/m$. The gluonic excitations (hybrids and glueballs) 
that form a gap  of order $\lQ$ with respect to the quarkonium state
can be integrated out  and the potentials follow
in an unambiguous and systematic way from the nonperturbative matching to pNRQCD.
Thus, we recover the quark model from pNRQCD. The US degrees of freedom in this case are 
not coloured gluons but US gluonic excitations between heavy quarks
 and pions. They  can be systematically included 
and  may eventually affect the leading potential picture. 
Let us consider for instance the singlet matching potential.
Disregarding the US corrections, we have the identification
$$V_s(r)  = \lim_{T\to\infty}{i\over  T    } 
\ln \langle W(r\times T)  \rangle .$$
US corrections to this formula are due to pions and US gluonic excitations between heavy quarks,  
and may be included in the same way as the effects due to US gluons have been included 
in the perturbative situation through Eq. (\ref{vsu0}).

The complete $1/m^2$ potential has been calculated along these lines \cite{m1}. 
There are many appealing and interesting features of this procedure. 
The matching calculation is performed in the way of the quantum mechanical 
perturbation theory on the QCD Hamiltonian (where perturbations are counted in orders of $1/m$
and not of $\alpha_s$) and only at a later stage the relation with the Wilson loop and field insertions 
is established. This allows us to have a control on the Fock states 
of the problem and on  the contributions coming from gluonic excitations  in the 
intermediate states. At the end, all the expressions are again given in terms of Wilson 
loops, which  can be evaluated on the lattice or in QCD vacuum models \cite{mod}.
The  potentials turn out to be naturally factorized in a hard part (the matching coefficients 
at the hard scales inherited by pNRQCD from NRQCD) and a low energy part (the Wilson loops
expressions). The power counting may turn out to be quite different from the perturbative 
(QED-like) situation.

\section{Conclusion and outlook}
I have shown that it is possible to construct systematically 
and within a controlled expansion an effective theory of QCD, which describes 
heavy quark bound states. All known perturbative and nonperturbative 
regimes (potential, nonpotential), are dynamically present in the theory, 
which  is equivalent  to QCD. I have presented many applications 
of pNRQCD in the situation $\lQ\ll mv$. In the situation $\lQ \simeq mv$, 
I have discussed  how pNRQCD allows us to systematically factorize  the nonperturbative 
heavy quark dynamics.

\section*{Acknowledgments}
I thank the organizers and especially M. Shifman and M. Voloshin for this interesting 
and very enjoyable workshop and for the local support; I acknowledge the 
Alexander Von Humboldt Foundation. I acknowledge 
the University of Milano for travel support. I thank A. Pineda, J. Soto and A. Vairo 
for collaboration on many topics presented here and A. Vairo for many discussions.

\end{document}